\documentclass[twocolumn,secnumarabic,amssymb, nobibnotes, aps, prd]{revtex4}
\usepackage[utf8]{inputenc}
\usepackage{amssymb,amsmath,graphicx,latexsym}
\usepackage{array}
\usepackage{bm}
\usepackage{color}
\usepackage{dcolumn}
\usepackage{multirow}
\usepackage{float}
\usepackage{ulem}
\usepackage{pdfpages}
\usepackage{threeparttable}
\usepackage{hyperref}

\newcommand{\dtabsize}{\small}
\newcommand{\clee}{C_{\ell}^{\rm EE}}
\newcommand{\xHI}{x_{\textsc{Hi~}}}

\begin{document}
\bibliographystyle{apsrev}
\title{Constraining the reionization history with CMB and spectroscopic observations}

\author{Wei-Ming Dai$^{1,3,4}$}
\email{daiwming@itp.ac.cn}

\author{Yin-Zhe Ma$^{3,4}$}
\email{Ma@ukzn.ac.za}

\author{Zong-Kuan Guo$^{1,2}$}
\email{guozk@itp.ac.cn}

\author{Rong-Gen Cai$^{1,2}$}
\email{cairg@itp.ac.cn}

\affiliation{
$^{1}$CAS Key Laboratory of Theoretical Physics, Institute of Theoretical Physics,
Chinese Academy of Sciences, P.O. Box 2735, Beijing 100190, China\\
$^{2}$School of Physical Sciences, University of Chinese Academy of Sciences,
No. 19A Yuquan Road, Beijing 100049, China\\
$^{3}$School of Chemistry and Physics, University of KwaZulu-Natal,
Westville Campus, Private Bag X54001, Durban, 4000, South Africa\\
$^{4}$NAOC-UKZN Computational Astrophysics Centre (NUCAC),
University of KwaZulu-Natal, Durban, 4000, South Africa}

\begin{abstract}
We investigate the constraints on the reionization history of the Universe from a joint analysis of the cosmic microwave background and neutral hydrogen fraction data. The $\tanh$ parametrization and principal component analysis methods are applied to the reionization history respectively. The commonly used $\tanh$ parametrization is oversimplistic when the neutral hydrogen fraction data are taken into account.
Using the principal component analysis method, the reconstructed reionization history is consistent with the neutral hydrogen fraction data.
With the principal component analysis method, we reconstruct the neutral hydrogen fraction at $z=9.75$ as $\xHI=0.69^{+0.30}_{-0.32}$ for $6<z<20$ range reconstruction, and $\xHI=0.76^{+0.22}_{-0.27}$ for $6<z<30$ range reconstruction. These results suggest that the Universe began to reionize at redshift no later than $z=10$ at a $95\%$ confidence level.
\end{abstract}

\maketitle

\section{Introduction}
\label{s:introduction}
The observation of the cosmic microwave background (CMB) radiation has provided state-of-the-art measurements on cosmological parameters.
Measurements from the Wilkinson Microwave Anisotropy Probe (WMAP) and Planck satellite have pinned down the precision of the reionization optical depth $\tau$ to an unprecedented level, which essentially constrains the reionization process.
There are two main effects of the reionization history on the CMB angular power spectra.
The first effect is the photon attenuation effect; i.e., the ionized electron rescatters the CMB photons which leads to a suppression of the acoustic peaks in the CMB angular power spectra. So the amplitude of the $C_\ell^{\rm TT}$ is proportional to $A_{\rm s} e^{-2\tau}$, where $A_{\rm s}$ is the the amplitude of the primordial curvature perturbations at the pivot scale $k_0=0.05\,{\rm Mpc}^{-1}$. Given the same ionized hydrogen fraction, it contributes more to the optical depth if the reionization process began earlier and lasted longer. The second effect is the reionization bump in the $C^{\rm TE}_{\ell}$ and $C^{\rm EE}_{\ell}$ power spectra, as the polarization is generated due to the quadrupole seen by electrons after reionization. The angular position of the bump is proportional to the square root of the redshift at which the reionization occurs,
while the amplitudes of $C^{\rm TE}_{\ell}$ and $C^{\rm EE}_{\ell}$ are proportional to $\tau$ and $\tau^{2}$ respectively~\cite{Zaldarriaga:1996ke,Hu:1996mn}.
Therefore, measurements of the large-scale polarization angular power spectra can strongly constrain the reionization history~\cite{Kaplinghat:2002vt,Hu:2003gh}.
The 9-year results of {\it WMAP} give an estimate of optical depth $\tau=0.089\pm0.014$~\cite{Bennett:2012zja}.
In the Planck 2015 analysis based on the temperature power spectra and low-$\ell$ polarization, the optical depth is found to be $\tau=0.078\pm0.019$~\cite{Ade:2015xua}.
Using the Planck-high frequency instrument E-mode polarization and temperature data, the Planck lollipop likelihood gives $\tau=0.058\pm0.012$~\cite{Adam:2016hgk}.

However, since the value of $\tau$ is an integration of free electron density,
the detailed process of reionization is still a mystery although we have a fairly precise measurement of $\tau$. A steplike instantaneous reionization model is proposed by Lewis~\cite{Lewis:2008wr} and used in the Planck 2013 and 2015 cosmological results.
Some variants of such a phenomenological model were considered to constrain the reionization history~\cite{Douspis:2015nca,Faisst:2014vra,Adam:2016hgk}.
A semianalytical reionization model is proposed based on the relevant physics governing these processes, such as the inhomogeneous intergalactic medium (IGM) density distribution, three different sources of ionizing photons, and radiative feedback~\cite{Choudhury:2004vs}.

All of the above models are built based on our current knowledge of the reionization.
If the ansatz of the reionization model is not accurate, the evaluated values of cosmological parameters may be biased.
Therefore, it is important and necessary to constrain it in a relatively model-independent way.
Hu and Holder~\cite{Hu:2003gh} proposed the principal component analysis (PCA) of the reionization history to quantify the information contained in the large-scale E-mode polarization.
This approach has been applied to both the simulated and real CMB data~\cite{Colombo:2008jr,Mortonson:2008rx}.
In our previous work, we applied such a PCA method for the reionization history to Planck 2015 data and found that the Universe is not completely reionized at redshift $z\gtrsim8.5$ at $95\%$ confidence level (C.L.)~\cite{Dai:2015dwa}.
The PCA method has been used to investigate the impacts of the reionization model on the estimates of cosmological parameters~\cite{Mortonson:2007tb,Mortonson:2009qv,Archidiacono:2010wp,Liu:2015gho,
Huang:2017huh,Heinrich:2016ojb}.
The estimated values of cosmological parameters such as the amplitude of the power spectrum of primordial scalar perturbations and neutrino masses are sensitive to the reionization history.

In addition, the evolution of the intergalactic Lyman-alpha (Ly$\alpha$) opacity measured in the spectra of quasars can provide valuable information on the reionization history~\cite{Gunn:1965hd}.
The recent measurements imply that the reionization of the IGM was nearly completed at redshift $z\approx 6$~\cite{Fan:2005es}.
The detection of complete Gunn-Peterson (GP) absorption troughs in the spectra of quasars at $z>6$ suggests that the neutral fraction of the IGM increases rapidly with redshift~\cite{Cen:2001us,Fan:2001vx,Lidz:2001yc,White:2003sc,Gnedin:2004nj}.
The rapid decline in the space density of Ly$\alpha$ emitting galaxies in the region $z=6 - 8$ implies a low-redshift reionization process~\cite{Choudhury:2014uba}.
But probing the high-redshift reionization history directly is still a big challenge.

In this paper, we apply two different methods to constrain the reionization history: the widely used $\tanh$ parametrization method proposed by Lewis~\cite{Lewis:2008wr} and the PCA approach proposed by Hu and Holder~\cite{Hu:2003gh}.
Using the Planck 2015 data combined with spectroscopic observations, we investigate the constraints on the reionization history and cosmological parameters.

This paper is organized as follows.
In Sec.~\ref{s:models}, we describe the $\tanh$ parametrization and PCA methods respectively.
In Sec.~\ref{s:xHI}, we list the current measurements of the neutral hydrogen fraction.
In Sec.~\ref{s:constraints}, we use the Planck 2015 data and the neutral hydrogen fraction data to put constraints on the reionization history.
Section~\ref{s:conclusion} is devoted to discussions and conclusions.

\section{Methods}
\label{s:models}

Throughout our analysis, we adopt a spatially flat $\Lambda$CDM model described by a set of cosmological parameters
$\{\Omega_{\rm b}h^2,\Omega_{\rm c}h^2,\theta_\mathrm{MC},A_{\rm s},n_{\rm s}\}$\,, where $\Omega_{\rm b}h^2$ and $\Omega_{\rm c}h^2$ are the physical baryon and cold dark matter densities relative to the critical density,
$\theta_\mathrm{MC}$ is an approximation to the ratio of the sound horizon to the angular diameter distance at the photon decoupling,
$A_{\rm s}$ is the amplitude defined as in Sec.~\ref{s:introduction}, and $n_{\rm s}$ is the spectral index of the primordial curvature perturbations at the pivot scale $k_0=0.05\,{\rm Mpc}^{-1}$.

\subsection{``$\tanh$'' function parametrization}
\label{s:parametrization}
The most widely used parametrization is a step like transition of the ionized hydrogen fraction $x_{\rm e}$, which is parametrized by the median redshift $z_{\rm re}$
and duration $\Delta z$ of the reionization.
A $\tanh$ function is utilized to fit the reionization history~\cite{Lewis:2008wr}:
\begin{equation}
x_{\rm e}^{\rm eff}(z)=f\ast x_{\rm e} (z)=\frac{f}{2}\left[1+\tanh\left(\frac{y(z_{\rm re})-y}{\Delta y}\right)\right]\,,
\label{eq:xe_inst}
\end{equation}
where $y=(1+z)^{3/2}$ and $\Delta y=3/2 (1+z)^{1/2}\Delta z$.
Since the first ionization energy ({24.6 \rm MeV}) of helium is not much higher than hydrogen (${13.6\, \rm MeV}$), it is usually assumed that helium first reionizes in the same way as hydrogen.
Ignoring the residual electron density from recombination, the efficient reionization fraction is $x_{\rm e}^{\rm eff}\equiv f\ast x_{\rm e}$. The factor $x_{\rm e}$ is the ratio between number densities of ionized hydrogen to the total hydrogen, and $f\ast x_{\rm e}$ is the number density ratio between free electrons and total hydrogen. Therefore the factor $f$ is $f=1+n_{\rm He}/n_{\rm H}$\,, where $n_{\rm He}$ and $n_{\rm H}$ are the number densities of helium and hydrogen respectively.
The typical value of $f$ is roughly 1.08 because the helium mass fraction is around $0.24$.
Additionally, we assume that hydrogen is fully ionized before the second helium ionization (the corresponding ionization energy is $54.4 \text{MeV}$). Meanwhile, the helium second reionizes at $z_{\text{re}}=3.5$ with the $\tanh$ parameters $f=n_{\text{He}}/n_{\text{H}}$ and $\Delta z = 0.5$.
The total efficient reionization fraction is the sum of contributions from hydrogen and helium.

It is argued that the hydrogen in the IGM could have been reionized twice~\cite{Wyithe:2002qu,Cen:2002zc}, although spectroscopic observations have given a hint that the IGM ionization is similar to a phase transition and the follow-up study revealed that double reionization requires extreme parameter choices~\cite{Furlanetto:2004nt}.
The simple parametrization described by Eq.~\eqref{eq:xe_inst} may bias the reionization history.
To eliminate the bias, we can define discrete ionization fractions in a series of small redshift bins, which correlate with each other in practice.

\subsection{Principal component analysis}
\label{s:pca}

The PCA method converts a set of correlated variables into a set of linear uncorrelated variables by an orthogonal transformation.
Most information is encoded in the principal components, which are picked out according to their corresponding eigenvalues.
Following Refs.~\cite{Mortonson:2007hq,Dai:2015dwa}, we consider a binned ionization fraction $x_{\rm e}(z_i)$, $i\in\{1,2,\ldots,N_z\}$, with redshift bins of width $\delta z=0.25$.
We take $z_{\rm min}=6$ and $z_{\rm max}=30$ with the definition $z_1=z_{\rm min}+\delta z$ and $z_{N_z}=z_{\rm max}-\delta z$ so that $N_z+1=(z_{\rm max}-z_{\rm min})/{\delta z}$.
The principal components of $x_{\rm e}(z_i)$ are the eigenfunctions of the following Fisher matrix $F_{i j}$,
\begin{equation}
F_{ij}=\sum_{\ell=2}^{\ell_{\rm max}}\left(\ell+\frac{1}{2}\right)
       \frac{\partial \ln \clee}{\partial x_{\rm e}(z_i)}
       \frac{\partial \ln \clee}{\partial x_{\rm e}(z_j)} \,,
\label{eq:fisher1}
\end{equation}
which describes the dependence of the polarization spectrum $\clee$ on the ionization fraction $x_{\rm e}(z_i)$.
The Fisher matrix $F_{ij}$ can be decomposed as
\begin{equation}
F_{ij}=(N_z+1)^{-2}\sum_{\mu=1}^{N_z}S_\mu(z_i)\sigma_\mu^{-2}S_{\mu}(z_j) \,,
\label{eq:fisher2}
\end{equation}
where $\sigma_\mu^2$ are the inverse eigenvalues and
$S_\mu(z)$ are the eigenfunctions that satisfy the orthogonality and completeness relations
\begin{eqnarray}
\int_{z_{\rm min}}^{z_{\rm max}} {\rm d}z S_\mu(z)S_\nu(z)&=&(z_{\rm max}-z_{\rm min})\delta_{\mu\nu} \,,
\label{eq:orthog} \\
\sum_{\mu=1}^{N_z} S_\mu(z_i)S_\mu(z_j)&=&(N_z+1)\delta_{ij} \,.
\label{eq:compl}
\end{eqnarray}
Then, the reionization history is represented in terms of the eigenfunctions as
\begin{equation}
x_{\rm e}(z)=x_{\rm e}^{\rm fid}(z)+\sum_\mu m_\mu S_\mu(z)\,.
\label{eq:xe_pca}
\end{equation}
The $x_{\rm e}^{\rm fid}$ is the fiducial value of the hydrogen reionization fraction, we set $x_{\rm e}^{\rm fid}=0.1$ in our fiducial models, and $m_\mu$ are the amplitudes of principal components.
Mortonson and Hu~\cite{Mortonson:2007hq} argued that $x_{\rm e}(z)$ is not necessarily bounded in between $0$ and $1$ at all redshifts and derived a necessary but not sufficient condition for physicality.
Nevertheless, in this paper, we simply assume that the selected principal components reconstruct the reionization history sufficiently well so that the bound is $x_{\rm e}(z)\in [0,1]$.
The reason is that we combine the neutral hydrogen fraction data listed in Sec.~\ref{s:xHI} with CMB to fit the cosmological parameters and our reionization model has no impact on the neutral hydrogen fraction data (physically, $\xHI \in [0,1]$, defined in Sec.~\ref{s:xHI}).

Since the reionization history is reconstructed with only the first few eigenvectors, there are some residual errors to be corrected by the rest of the eigenvectors.
In practice, we can regard these truncation errors as systematic errors introduced by the PCA method and make a rough estimate.
We emphasize that any truncation of principal component decomposition provides the least squares approximation of the real reionization history as the eigenfunctions satisfy the orthogonality and completeness relations.
Heinrich {\it et al}. quantitatively demonstrated that the first five eigenvectors form a complete representation of the observable impact on $\clee$ of any given reionization history, but the representation is not complete in the ionization history itself~\cite{Heinrich:2016ojb}.
The present data are incapable of putting a strict limit on the reionization history without any physical hypothesis, which means the possible function space [$z\mapsto x_{\rm e}(z)$] remains undetermined under this circumstance.
Moreover, the principal component decomposition is carried out based on the cosmic variance limited $\clee$ power spectrum instead of the real observation, which could loosen the constraint on the reionization history.
In what follows, we illustrate the character of PCA reconstructed reionization history and estimate the systematic errors via two types of general reionization models.

\begin{figure}
\begin{center}
\includegraphics[width=3.5in,height=2.5in]{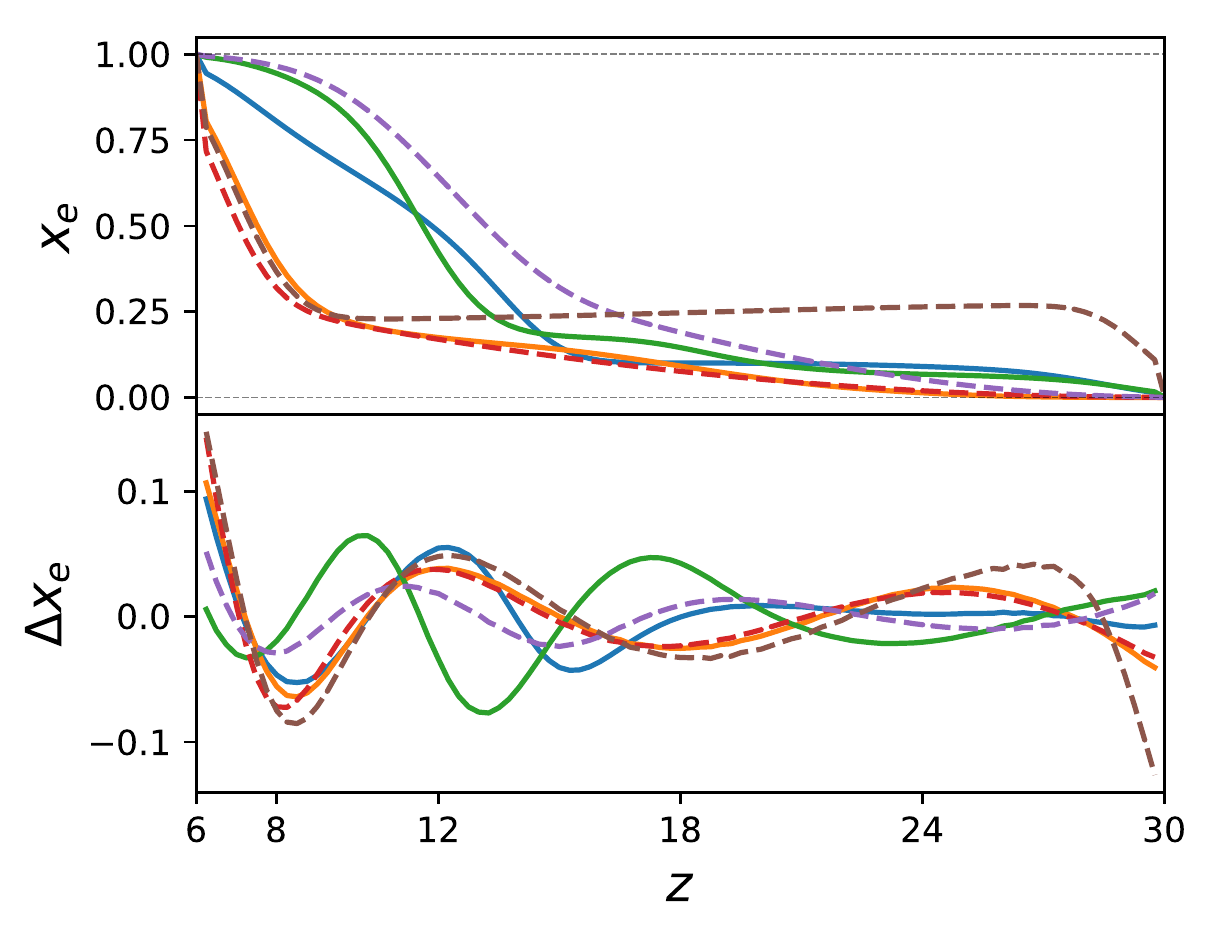}
\caption{Randomly sampled reionization history (upper panel) and the corresponding systematic errors of principal component decomposition (lower panel). Functions of $x_e(z)$ are constructed by connecting the randomly sampled knots and the end points with PCHIP. The solid lines
are type (a) curves, while the dashed lines are type (b) curves. All of them are smoothed by Gaussian function with
$\sigma=4\delta z=1.0$. Curves of $\Delta x_e(z)$ are clustered around $\Delta x_e(z)=0$.}
\label{fig:exp_eor_err}
\end{center}
\end{figure}

To get rid of the unphysical curves,
we generate the samples under the additional condition $0.03<\tau<0.13$, which is around
$3\sigma$ width of Planck constraints.
Figure~\ref{fig:exp_eor_err} illustrates the cases of randomly sampled reionization history.
Each curve connects the end points and randomly sampled knots, while the interpolation function is a piecewise cubic Hermite interpolating polynomial (PCHIP).
This approach makes sure that $x_e(z)$ is bounded in between 0 and 1.
We consider two types of curves: (a) monotonically decreasing curves interpolated between the end points ($z=6.0$ and $z=30.0$) and five randomly sampled knots with PCHIP and
(b) nonmonotonic curves interpolated between the end points and two randomly sampled knots with PCHIP.
All curves are smoothed by a Gaussian function.
Then, we project $x_e(z)$ onto the eigenvectors and get the coefficients
\begin{equation}
m_\mu=\frac{\sum_i [x_e(z_i)-x_e^{\text{fid}}(z_i)] S_\mu(z_i)}{\sum_i S_\mu(z_i) S_\mu(z_i)} \,.
\end{equation}
With Eq.~\eqref{eq:xe_pca}, the PCA reconstruction is easy and straightforward.
$\Delta x_e(z)$ is defined as the difference between PCA reconstructed and the true form of $x_e(z)$, which is sensitive to the reionization history.

\begin{figure}
\begin{center}
\includegraphics[width=3.5in,height=2.5in]{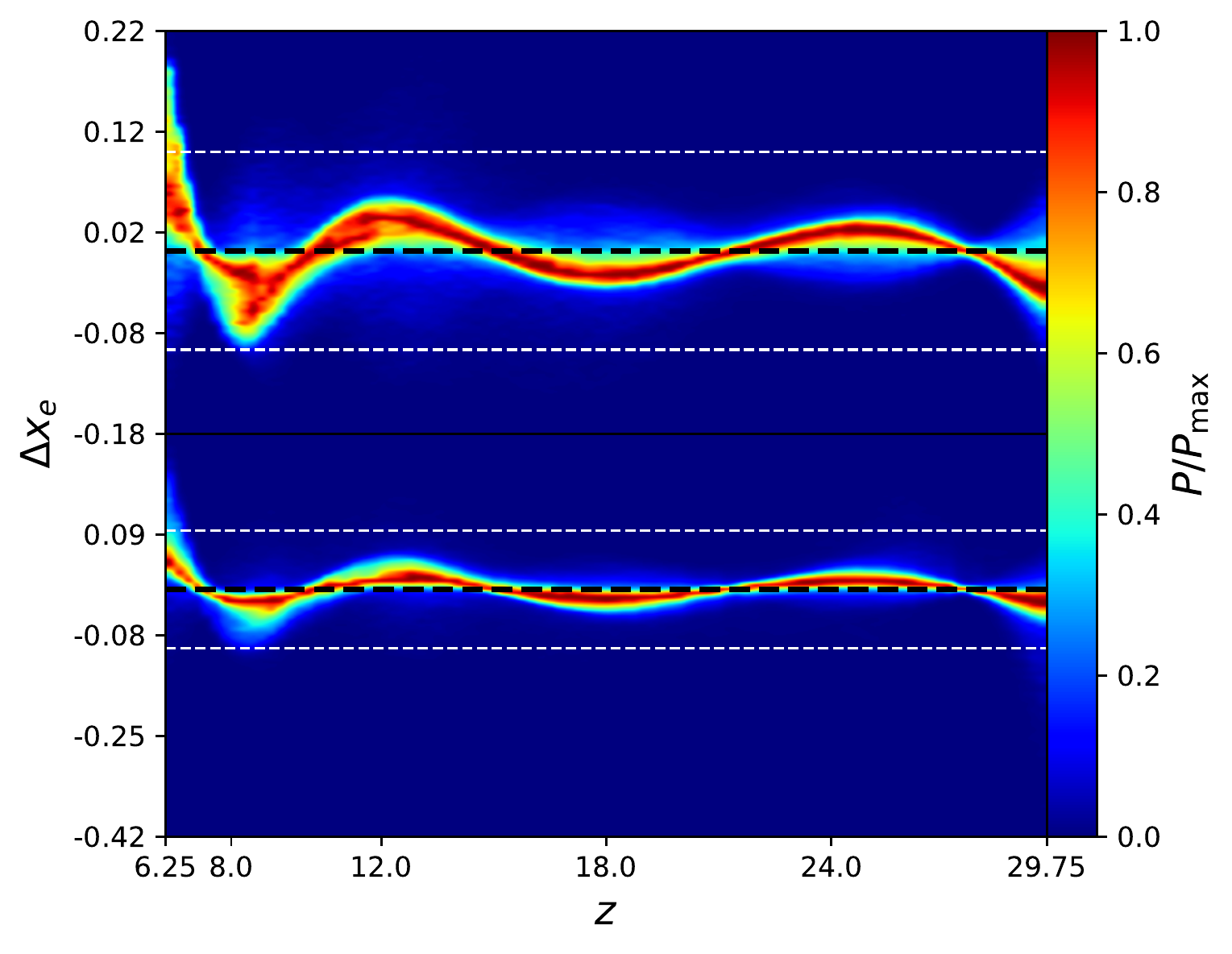}
\caption{Distribution of the systematic errors of the PCA method. The upper panel illustrates the type (a) curves, while the lower panel illustrates the type (b) curves. The black and white dashed horizontal lines show a band with width $\Delta x_e(z) \le 0.1$, centered at $\Delta x_e(z)=0$.}
\label{fig:err_img}
\end{center}
\end{figure}

\begin{figure}
\begin{center}
\includegraphics[width=3.5in,height=2.5in]{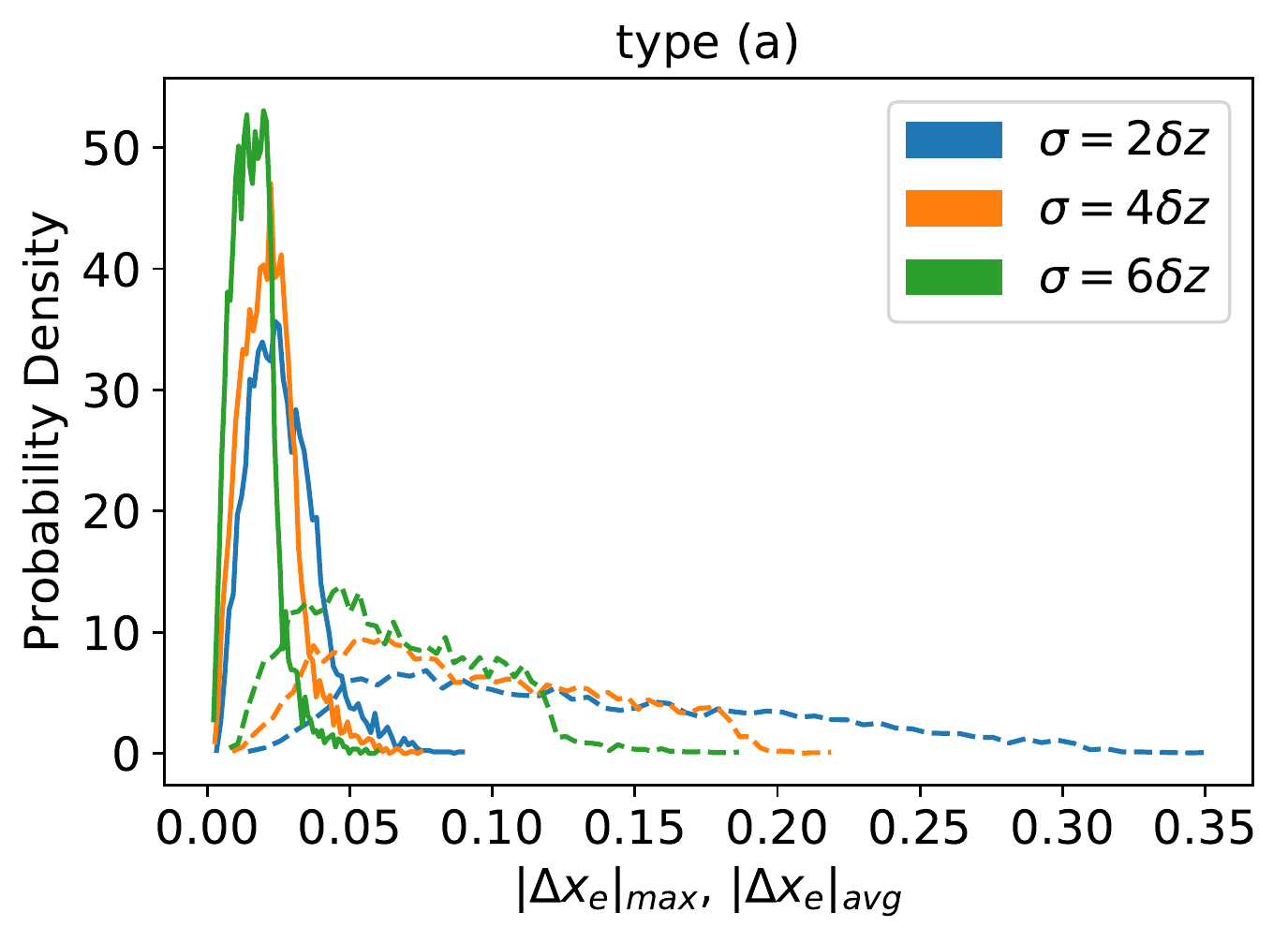}
\includegraphics[width=3.5in,height=2.5in]{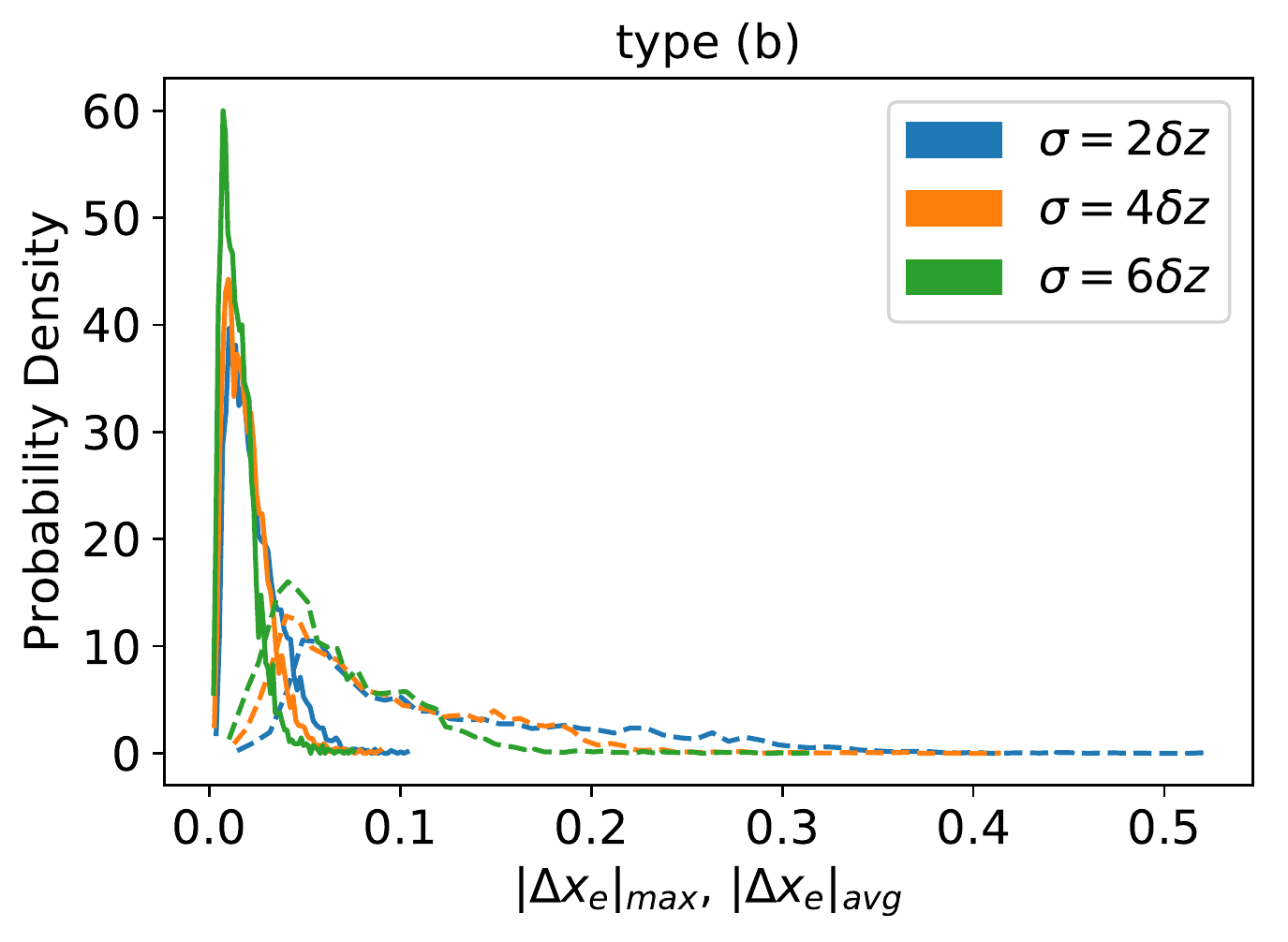}
\caption{Estimation on the systematic errors of the PCA method. The solid curves show the probability density of the average of $|\Delta x_e|$, while the dashed curves show the probability density of the maximal value of each reionization instance. Three smoothing scales of the reionization curves are plotted: $\sigma=2\delta z= 0.5$, $\sigma=4\delta z= 1.0$ and $\sigma=6\delta z= 1.5$.}
\label{fig:mid_max}
\end{center}
\end{figure}

Figure~\ref{fig:err_img} shows the distribution of $\Delta x_e(z)$ within the redshift range $6.25\le z\le 29.75$, assuming that each possible reionization history is of equal probability as in Fig.~\ref{fig:exp_eor_err}.
It suggests that the PCA method is also likely to bias the reionization history, although it is independent of any physical hypothesis.
But statistically, the bias is expected to be small for a general reionization curve.
To estimate the possible systematic errors,
we compute the probability density of $|\Delta x_e|_{\max}$, the maximal value of systematic errors of each reionization instance,
as well as the probability density of $|\Delta x_e|_{\text{avg}}$, the average of systematic errors of each reionization instance,
defined as $|\Delta x_e|_\text{avg}=\int_{z_{\min}}^{z_{\max}}|\Delta x_e(z)| dz /(z_{\max}-z_{\min})$.
Figure~\ref{fig:mid_max} shows the probability density of the maximal errors $|\Delta x_e|_{\max}$ and the average errors $|\Delta x_e|_{\text{avg}}$.
$|\Delta x_e|_{\text{avg}}$ is approximately bounded between $0$ and $0.05$ for all cases.
In the case that the Gaussian smoothing scale is $\sigma=1.0$, the boundary of the maximal error is $|\Delta x_e|_{\max}\lesssim 0.2$.
The distribution can be sharpened by increasing the smoothing scale.
That means the PCA method keeps the overall feature, but is incapable of catching the local property.
The PCA method is applicable, since we investigate the reionization history in a large redshift range and
probably lose the local details.

In what follows, ``instant" denotes the parametrization method for the reionization history~[Eq.\eqref{eq:xe_inst}] and ``PCA" denotes the PCA method~(Eq.\eqref{eq:xe_pca}).
The former is described by the median redshift $z_{\rm re}$ and reionization duration $\Delta z$,
while the latter is described by five parameters $m_\mu$, $\mu =1,...,5$.
In our analysis we use the publicly available CosmoMC package to explore the parameter space by means of the Markov chain Monte Carlo (MCMC) technique~\cite{Lewis:2002ah}.
We modify the Boltzmann {\sc camb} code~\cite{Lewis:1999bs} to appropriately incorporate the reionization history.
The reionization parameters and other cosmological parameters are evaluated by performing global fitting in Sec.~\ref{s:constraints}.

\section{Data}
\label{s:xHI}

\begin{table*}
\renewcommand\arraystretch{1.2}
\begin{center}
\caption{The current constraints on the neutral hydrogen fraction $\xHI$ from different observations, ranging from low to high redshifts. ``LAEs" means Ly$\alpha$ emitters, i.e. Ly$\alpha$ emitting galaxies.}
\begin{threeparttable}
\begin{tabular}{>{\dtabsize}c|>{\dtabsize}c|>{\dtabsize}c|>{\dtabsize}m{0.4\columnwidth}<{\centering}|>{\dtabsize}m{0.4\columnwidth}<{\centering}|>{\dtabsize}c|>{\dtabsize}c|>{\dtabsize}c}
\hline
\hline
Redshift $z$  &$\xHI$ data  &C.L.  &Technique  &Observation  &Ref.  &Year   &Dataset \\
\hline
5.03  &$\xHI=(5.5\times10^{-5})^{+1.42\times10^{-5}}_{-1.65\times10^{-5}}$  &$1\sigma$  &GP optical depth of QSOs  &SDSS  &\cite{Fan:2005es}  &2006 &Full \\
5.25  &$\xHI=(6.7\times10^{-5})^{+2.07\times10^{-5}}_{-2.44\times10^{-5}}$  &$1\sigma$  &                          &      &                                 \cite{Bouwens:2015vha} &     &     \\
5.45  &$\xHI=(6.6\times10^{-5})^{+2.47\times10^{-5}}_{-3.01\times10^{-5}}$  &$1\sigma$  &                          &      &                                   &     &     \\
5.65  &$\xHI=(8.8\times10^{-5})^{+3.65\times10^{-5}}_{-4.60\times10^{-5}}$  &$1\sigma$  &                          &      &                                   &     &     \\
5.85  &$\xHI=(1.3\times10^{-4})^{+4.08\times10^{-5}}_{-4.90\times10^{-5}}$  &$1\sigma$  &                          &      &                                   &     &     \\
6.10  &$\xHI=(4.3\times10^{-4})\pm({3.0\times10^{-4}})$                     &$1\sigma$  &                          &      &                                   &     &     \\
\hline
5.3  &$\log_{10}\xHI=-4.4^{+0.84}_{-0.90}$  &$1\sigma$  &QSO dark gap statistics  &SDSS  &\cite{Gallerani:2007ty}  &2008  &Full \\
5.6  &$\log_{10}\xHI=-4.2^{+0.84}_{-1.0}$   &$1\sigma$  &                         &      &                         &      &     \\
\hline
5.6  &$\xHI<0.04+0.05$  &$1\sigma$  &Counts of dark Lyman-alpha pixels  &Keck II telescopes  &\cite{McGreer:2014qwa}  &2015  &Full \\
5.9  &$\xHI<0.06+0.05$  &$1\sigma$  &                                   &                    &                        &      &     \\
\hline
6.247 &$\xHI\gtrsim0.14$  &$2\sigma$  & QSO damping wing J1623+3112  &SDSS  &\cite{Schroeder:2012uy}  &2013  &Full/ext \\
6.308 &$\xHI\gtrsim0.11$  &$2\sigma$  &   J1030+0524          &      &                         &      &         \\
6.4189 &$\xHI\gtrsim0.14$  &$2\sigma$  &   J1148+5251                  &      &                         &      &         \\
\hline
6.3  &$\xHI=0.0\pm0.17\pm0.60$  &$1\sigma, 2\sigma$  &Ly$\alpha$ damping wing of GRB 050904  &Subaru Telescope  &\cite{Totani:2005ng} &2006  &Full/ext \\
\hline
6.3  &$\xHI=(6.4\pm0.3)\times10^{-5}$  &$1\sigma$  &GRB 050914 spectra  &Swift satellite  &\cite{Gallerani:2007nb}  &2008  &Not applicable  \\
\hline
6.5  &$\xHI\lesssim0.3$  &N/A  &LAEs  &Large-Area Lyman Alpha survey  &\cite{Malhotra:2004ef}  &2004  &Not applicable  \\
\hline
6.5  &$0\lesssim \xHI\lesssim0.45$  &N/A  &17 LAEs  &Subaru Deep Field and Keck  &\cite{Kashikawa:2006pb}  &2006  &Not applicable  \\
\hline
6.6  &$\xHI=0.3\pm0.2$  &$1\sigma$  &2,354 LAEs  &Subaru/Hyper Suprime-Cam survey  &\cite{2017arXiv170501222K}  &2017  &Full/ext \\
\hline
6.6  &$\xHI\lesssim 0.2\pm 0.2$  &N/A  &207LAEs  &subaru/XMM-Newton Deep Survey field  &\cite{Ouchi:2010wd}  &2010  &Not applicable   \\
\hline
6.6  &$\xHI< 0.5$  &$2\sigma$  & Clustering of 58 LAEs  &Subaru Deep Field  &\cite{McQuinn:2007dy}  &2007  &Full/ext  \\
\hline
6.6 &$\xHI\simeq0.24 - 0.36$  &N/A  &Model and observed Ly$\alpha$ luminosity function  &Subaru Deep Field  &\cite{Ota:2007nx}  &2008  &Not applicable  \\
7.0 &$\xHI\simeq0.24 - 0.36$  &N/A  &                                                   &                   &                   &      &  \\
\hline
6.9  &$\xHI=0.4 - 0.6$  &N/A  &LAEs  &DECam/Blanco telescope  &\cite{Zheng:2017rvy}  &2017  &Not applicable  \\
\hline
7.0  &$\xHI=0.39^{+0.08}_{-0.09}$  &$1\sigma$  &LAEs  &Keck MOSFIRE spectrograph  &\cite{Schenker:2014tda}  &2014  &Full/ext \\
8.0  &$\xHI>0.64$                  &$1\sigma$  &      &                           &                         &      &         \\
\hline
7.0  &$\xHI>0.4$\tnote{a}          &$1\sigma$  &Ly$\alpha$ fraction evolution  &Numerical Simulation  &\cite{Mesinger:2014mqa}  &2015  &Full/ext \\
\hline
7.0  &$\xHI\sim0.5$   &N/A  &Prevalence of Ly$\alpha$ Emission in Galaxies  &Vary Large Telescope  &\cite{Caruana:2013qua}  &2014  &Not applicable  \\
\hline
7.0  &$\xHI\sim0.6 - 0.9$  &N/A  &Prevalence of Ly$\alpha$ Emission in Galaxies  &Keck Telescope  &\cite{Ono:2011ew}  &2012  &Not applicable  \\
\hline
7.0  &$\xHI\geq0.51$  &N/A  &Prevalence of Ly$\alpha$ Emission in Galaxies  &Vary Large Telescope  &\cite{Pentericci:2014nia}  &2014  &Not applicable \\
\hline
7.0  &$\xHI\lesssim0.5$  &$1\sigma$  &Clustering of LAEs  &Subaru Hyper Suprime-Cam  &\cite{Sobacchi:2015gpa}  &2015  &Full/ext  \\
\hline
7.085  &$\xHI\gtrsim0.1$  &N/A  &Quasar ULAS J1120 + 0641  &UKIRT Infrared Deep Sky Survey  &\cite{Mortlock:2011va}  &2011  &Not applicable  \\
\hline
7.085  &$\xHI=0.40^{+0.21+0.41}_{-0.19-0.32}$  &$1\sigma, 2\sigma$  &ULAS J1120 + 0641 damping wing  &Magellan/Baade telescope  &\cite{Greig:2016vpu}  &2017  &Full/ext \\
\hline
8.0  &$\xHI\gtrsim0.3$  &N/A  &Prevalence of Ly$\alpha$ Emission in Galaxies  &Keck Telescope  &\cite{Tilvi:2014oia}  &2014  &Not applicable  \\
\hline
\end{tabular}
\label{tab:xHI}
\begin{tablenotes}
\item[a] Converted from ionized fraction. These data are derived from numerical simulation rather than observation.
\end{tablenotes}
\end{threeparttable}
\end{center}
\end{table*}

\begin{figure*}
\centerline{
\includegraphics[width=3.5in,height=1.7in]{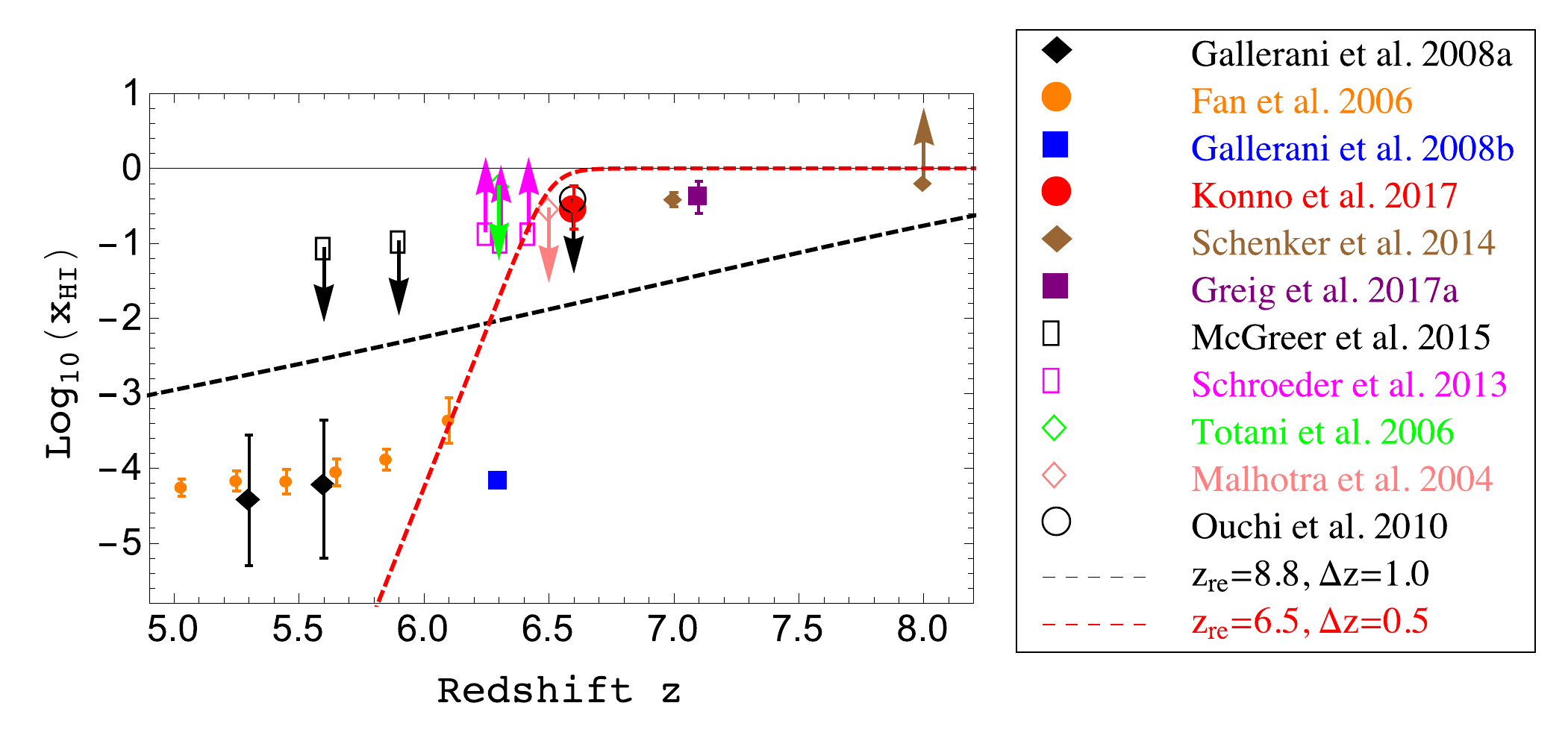}
\includegraphics[width=3.5in,height=1.7in]{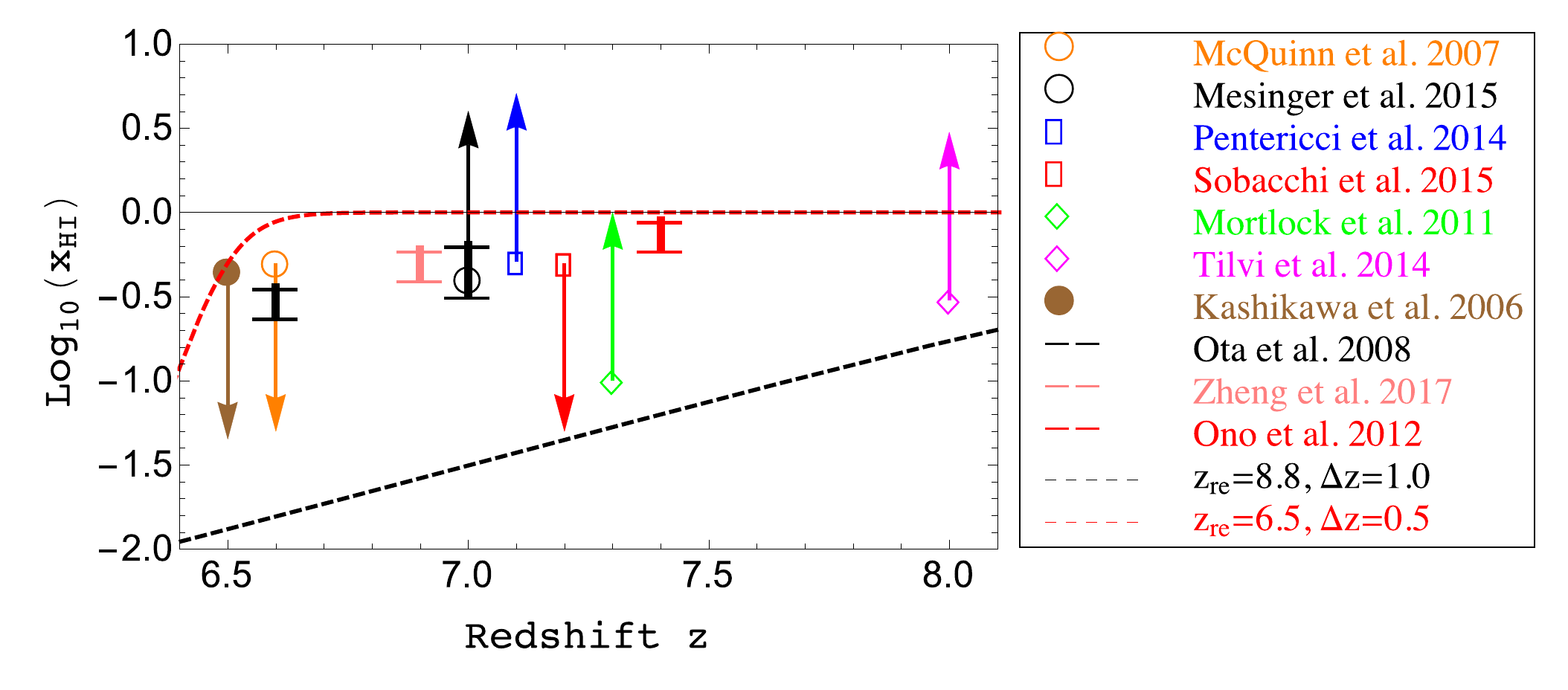}}
\caption{The state-of-the-art measurement on $\xHI (z)$, taken from Table~\ref{tab:xHI}. The black and red dashed lines are two examples of the ``$\tanh$” model which cannot fit the data very well.}
\label{fig:xHI}
\end{figure*}

We list current constraints on the volume-averaged neutral hydrogen fraction in Table~\ref{tab:xHI}.
Table~\ref{tab:xHI} summarizes the constraints on the neutral hydrogen fraction or free electron fraction over the redshift range $z=5$--$8$ which were derived from 2006 to 2017. These constraints can be summarized into four categories.
\begin{itemize}

\item Quasar/GRB Ly$\alpha$ absorption line systems~\cite{Fan:2005es,Gallerani:2007ty,McGreer:2014qwa,Schroeder:2012uy,Totani:2005ng,Gallerani:2007nb,Mortlock:2011va}.

\begin{enumerate}

\item Fan {\it et al}.~\cite{Fan:2005es} used the GP optical depth and \textsc{Hii~} region size measurements around luminous quasars to measure that the reionization process finishes between $z=5.9$ and $z=6.5$.

\item Gallerani {\it et al}.~\cite{Gallerani:2007ty} and McGreer {\it et al}.~\cite{McGreer:2014qwa} used quasar-stellar object (QSO) dark gap statistics and measured the fraction of neutral hydrogen to be very low at redshift $z \sim 5.6$.

\item Schroeder {\it et al}.~\cite{Schroeder:2012uy} used the GP damping wing of the spectra of three quasars (SDSS J1148+5251 ($z=6.4189$), J1030+0524 ($z=6.308$) and J1623+3112 ($z = 6.247$)), to constrain the neutral hydrogen fraction, $\xHI=1-x_{\rm e}$, and found the lower limit of $\xHI$ at $z\sim 6.2$--$6.4$.

\item Totani {\it et al}.~\cite{Totani:2005ng} used the Ly$\alpha$ damping wing of GRB 050914 ($z=6.3$) spectra to obtain the column density of $\textsc{Hi}$, and derived the upper limit of $\xHI$ to be $\xHI<0.17$ and $0.60$ at 68\% and 95\% C.L. respectively.

\item Gallerani {\it et al}.~\cite{Gallerani:2007nb} used the dark portions (gaps) in GRB 050904 absorption spectra to derive the neutral hydrogen fraction $\xHI=(6.4 \pm 0.3)\times 10^{-5}$ at $z=6.29$.

\item Mortlock {\it et al}.~\cite{Mortlock:2011va} reported a quasar (ULAS J112001.48+064124.3) at $z=7.085$, and used the Ly$\alpha$ damping wing profile to obtain that the neutral fraction of the intergalactic medium in front of ULAS J1120+0641 exceeded $0.1$. Using the same quasar, Greig {\it et al}.~\cite{Greig:2016vpu} accounted for uncertainties of the intrinsic QSO emission spectrum and the distribution of cosmic $\textsc{Hi}$ patches during the epoch of reionization (EoR) from simulation and reported that the EoR is not yet complete by $z= 7.1$, with the
volume-weighted IGM neutral fraction constrained to be $\xHI=0.40^{+0.21+0.41}_{-0.19-0.32}$ at $1\sigma$ and $2\sigma$ C.L.

\end{enumerate}

\item The number density and clustering of Ly$\alpha$ emitting galaxies~\cite{Malhotra:2004ef,Kashikawa:2006pb,2017arXiv170501222K,Ouchi:2010wd,Ota:2007nx,Zheng:2017rvy,Schenker:2014tda}. This type of observation is to use Ly$\alpha$ emitting galaxies to measure the Ly$\alpha$ luminosity functions and then by comparing the Ly$\alpha$ luminosity function measurements with reionization models, one can derive the neutral hydrogen fraction of the intergalactic medium $\xHI$. Such studies give the measurement of $\xHI$ in the redshift range of $6.5$ to $8.0$.

\item Gravitational clustering of Ly$\alpha$ emitters~\cite{McQuinn:2007dy,Sobacchi:2015gpa}. As shown in~\cite{McQuinn:2007dy,Sobacchi:2015gpa}, reionization increases the measured clustering of emitters, which can be computed observationally. By comparing the observational clustering of emitters with the results using radiative transfer simulations, McQuinn {\it et al}.~\cite{McQuinn:2007dy} and Sobacchi and Mesinger.~\cite{Sobacchi:2015gpa} obtained the upper limit of $\xHI \lesssim 0.5$ at $z=6.6$ and $7.0$ respectively.

\item Prevalence of Ly$\alpha$ emission in galaxies at redshift $6$--$8$~\cite{Caruana:2013qua,Ono:2011ew,Pentericci:2014nia,Tilvi:2014oia}. This class of observation is to assume that Ly$\alpha$ emission is prevalent in star-forming galaxies at $z\sim 6.5$--$8$, which is a simple extrapolation of the observed prevalence at $z \sim 4$--$6$. Then any departure from these trends is due to an
increasingly neutral IGM at $z \sim 7$--$8$. Therefore one can use this technique
to quantify the filling factor of ionized hydrogen ($Q_{\textsc{Hii}}$) at $z\sim 6.5$--$8$. Then one can convert this factor to IGM fractional neutral hydrogen density $\xHI$.

\end{itemize}

As marked in the last column of Table~\ref{tab:xHI}, we divide the $\xHI$ data into different datasets.
Only the data with C.L. are used in our analysis,
while the others are plotted in figures for comparison.
The error bar is conservatively estimated if it is not given explicitly.
For example, since a lower limit is given in Ref.~\cite{Mesinger:2014mqa} we assume that the mean value is $\xHI=1$ , and the mean value is $\xHI=0$ for the upper limit given in Ref.~\cite{Sobacchi:2015gpa}.
Because the limit derived in Ref.~\cite{Gallerani:2007nb} is much tighter than the others,
we do not use these data in our analysis.
In the PCA model, we assume that the reionized fraction $x_{\rm e}$ is exact unity at $z\leq6.0$.
The dataset of $\xHI$ used to constrain the reionization history in the PCA model is denoted by ``ext" in Table~\ref{tab:xHI}.
All  data given with confidence level can be used in the instant model, which is denoted by ``full."
Based on the common instant reionization assumption, we obtain a $\tanh$ model of $\xHI$ increasing with $z$.
The $\tanh$ model is intuitively compared with $\xHI$ data in Fig.~\ref{fig:xHI} with ($z_{\rm re}=8.8$, $\Delta z=1.0$) and ($z_{\rm re}=6.5$, $\Delta z=0.5$). For these two selective values, the $\tanh$ model cannot match the data very well.

\section{Results}
\label{s:constraints}

In our analysis, besides the neutral hydrogen fraction data,
we use Planck 2015 likelihood code and data,
including the Planck low-$\ell$ likelihood at multipoles $2\le \ell \le 29$ and high-$\ell$ PlikTT likelihood at multipoles $\ell\ge 30$ based on pseudo-$C_{\ell}$ estimators.
The low-$\ell$ likelihood uses the foreground-cleaned LFI 70 GHz polarization maps together with the temperature map obtained from the Planck 30 to 353 GHz channels by the Commander component separation algorithm over 94\% of the sky.
The high-$\ell$ PlikTT likelihood uses 100, 143, and 217 GHz cross-half-mission temperature spectra,
avoiding the Galactic plane as well as the brightest point sources and the regions where the CO emission is the strongest.
Hereafter, ``Planck 2015'' denotes the combination of the PlikTT temperature likelihood and the low-$\ell$ temperature-polarization likelihood.

\begin{figure}
\begin{center}
\includegraphics[width=3.5in,height=2.5in]{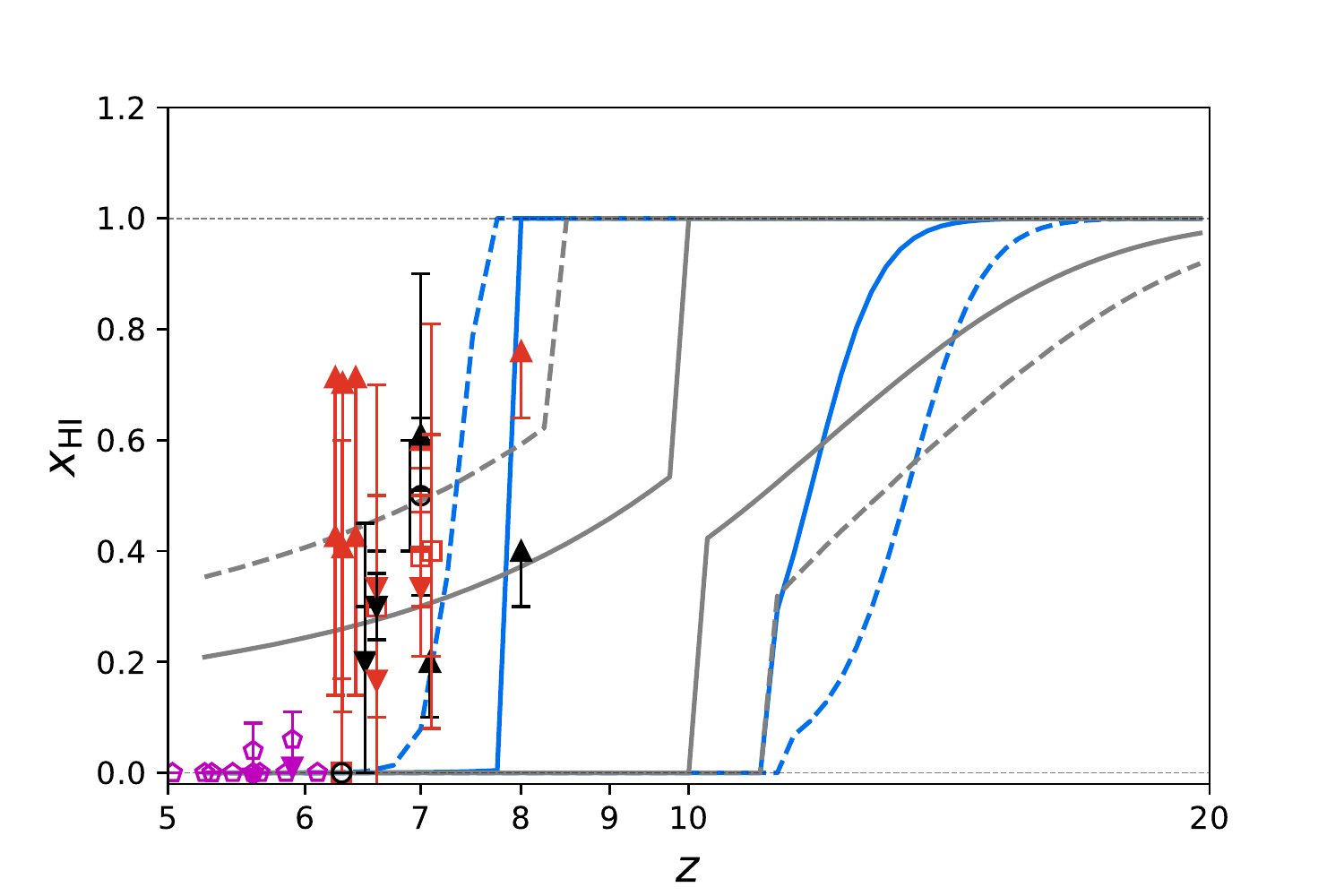}
\caption{Limits on $\xHI$ at redshift ranging from 5 to 20 in the instant reionization model, with 68\% (solid) and 95\% (dashed) confidence regions, derived from Planck 2015 + full data (blue) and Planck 2015 (gray) respectively. The red and magenta points as well as error bars belong to the full dataset as marked in Table~\ref{tab:xHI}, while the black points and error bars are not applicable in our analysis and just plotted for visual comparison}
\label{fig:inst_zre_delta_Full}
\end{center}
\end{figure}

\begin{figure*}
\begin{center}
\includegraphics[width=6.0in,height=4.0in]{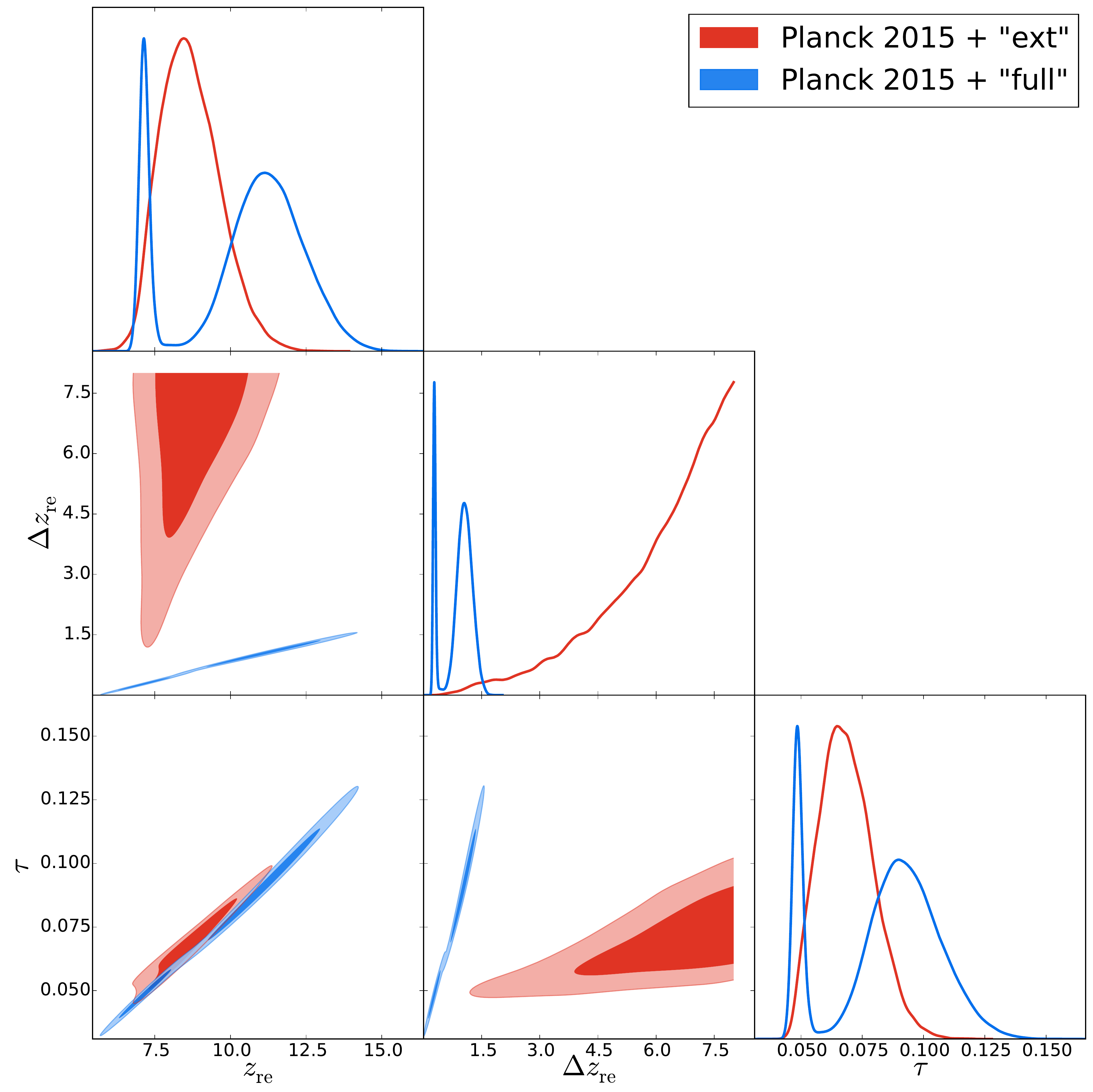}
\caption{Marginalized 2D contours ($68\%$ and $95\%$ C.L.) and posterior distributions for parameters of the instant reionization model, derived from Planck 2015 + full (blue) and Planck 2015 + ext (red) respectively.}
\label{fig:joint_inst_tri}
\end{center}
\end{figure*}

We constrain the instant model of the EoR with Planck 2015 data and the full $\xHI$ data.
We reconstruct the EoR during the redshift interval $5<z<20$ in Fig.~\ref{fig:inst_zre_delta_Full}.
The transition occurs at the redshift ranging from $z\sim 8$ to $14$.
The reconstructed figure is not fully consistent with the $\xHI$ data.
Meanwhile, in Fig.~\ref{fig:joint_inst_tri}, we see that the posterior distributions of $\tau$, $z_{\rm re}$ and $\Delta z$ are bimodal.
This means the instant model may bias the EoR.
The 2D contours derived from Planck 2015 + ext are also plotted in Fig.~\ref{fig:joint_inst_tri}.
There are no $\xHI$ data at redshift $z<6$ in the ext dataset, which means we remove the limit that the Universe is fully ionized at $z\sim 6$ in this model.
But the estimated median redshift and duration of reionization are $z_{\rm re}\sim 8$ and $\Delta z\gtrsim 8$.
This gives an unphysical result that the Universe is still not fully ionized today.

\begin{figure}
\begin{center}
\includegraphics[width=3.5in,height=2.5in]{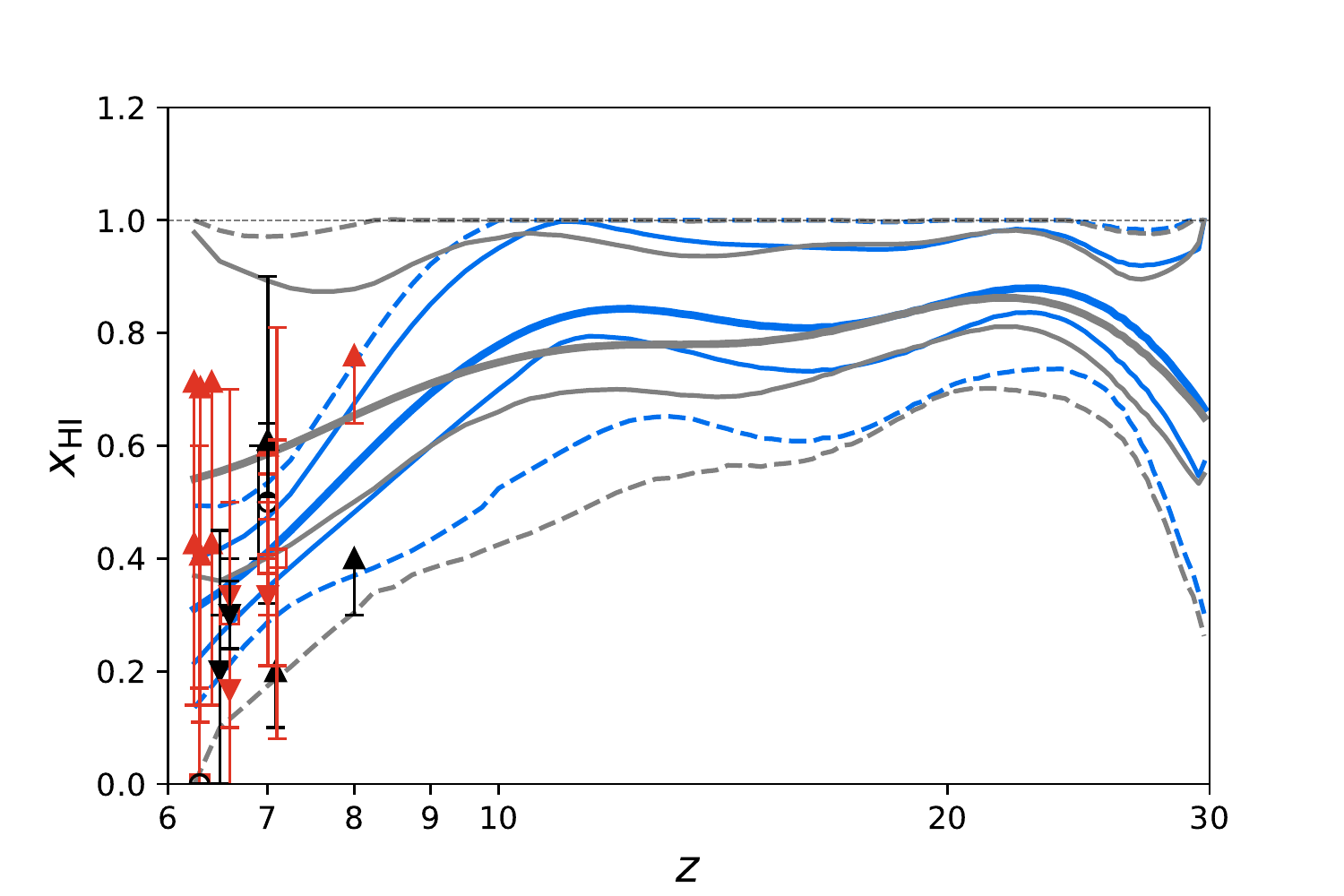}
\caption{Limits on $\xHI$ at redshift ranging from 6 to 30 in the PCA model, with 68\% (solid) and 95\% (dashed) confidence regions and mean values (thick solid), derived from Planck 2015 + ext data (blue) and Planck 2015 (gray) respectively. The red points as well as error bars belong to the ext dataset as marked in Table~\ref{tab:xHI}, while the black points and error bars are not applicable in our analysis and just plotted for visual comparison.}
\label{fig:noExtreme_Full}
\end{center}
\end{figure}

We constrain the PCA model of EoR with Planck 2015 + ext data, with a redshift interval of $6<z<30$.
As plotted in Fig.~\ref{fig:noExtreme_Full}, the reconstructed $\xHI(z)$ function covers the $\xHI$ data.
The error bar of the optical depth $\tau$ is smaller than in the instant model as shown in Table~\ref{tab:parameters}.
Comparing the confidence regions derived from Planck 2015 + ext data (blue) and Planck 2015 (gray) in Fig.~\ref{fig:noExtreme_Full},
we see that constraints on $x_{\rm e}$ between $z\sim 6$ and $z\sim 10$ are strengthened with the help of $\xHI$ data.
But the additional data do not have a significant impact on the high-redshift EoR.

\begin{figure}
\begin{center}
\includegraphics[width=3.5in,height=2.5in]{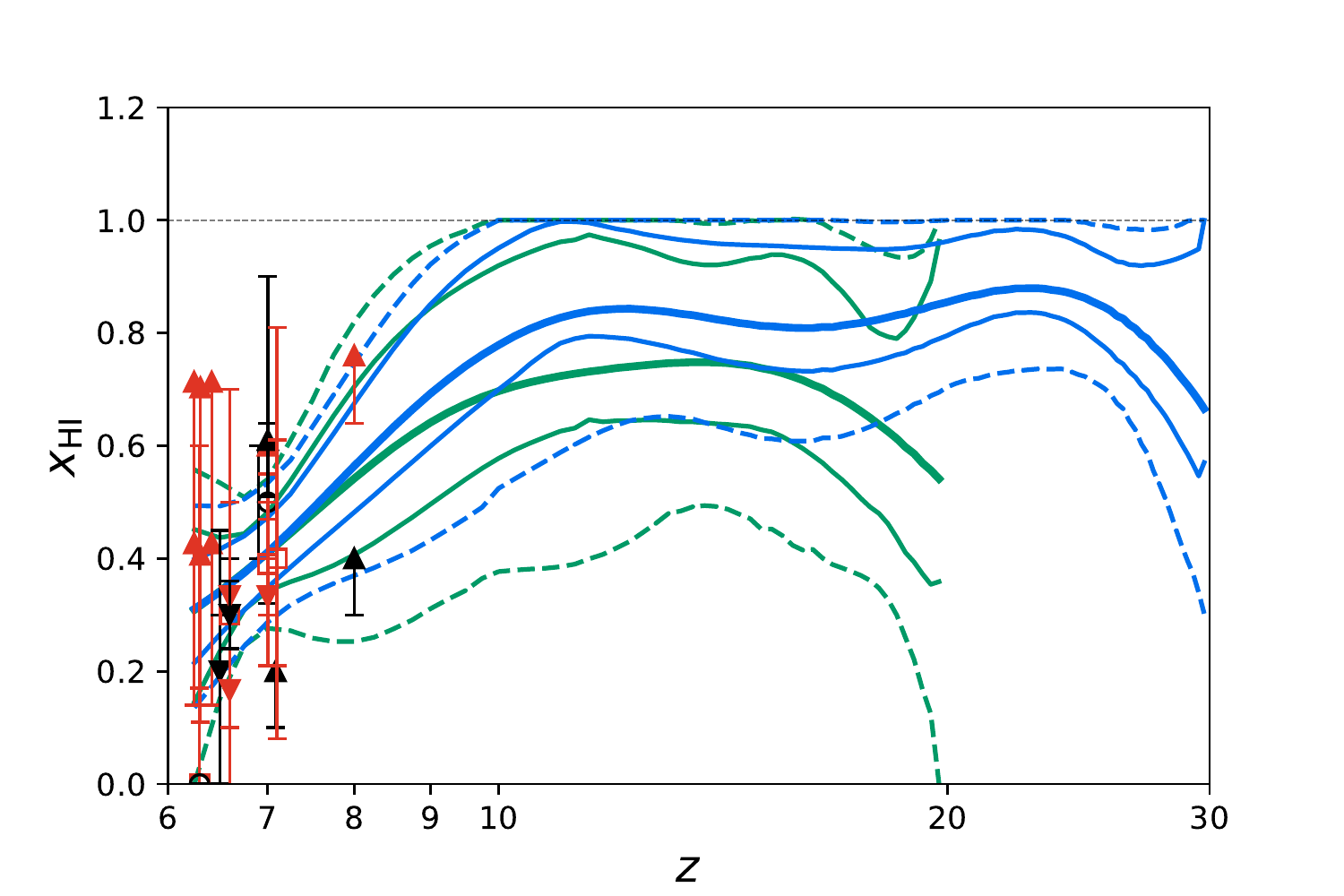}
\caption{Limits on $\xHI$ at redshift ranging from 6 to 20 (green) and 6 to 30 (blue) in the PCA model, with 68\% (solid) and 95\% (dashed) confidence regions and mean values (thick solid), derived from Planck 2015 + ext data. The red points as well as error bars belong to the ext dataset as marked in Table~\ref{tab:xHI}, while the black points and error bars are not applicable in our analysis and just plotted for visual comparison.}
\label{fig:noExtreme_Full_6to20}
\end{center}
\end{figure}

We also limit the range of reconstruction to be $6<z<20$ in the PCA model, and we obtain that the mean value of $\tau$ decreases by about $1\sigma$ C.L. The reconstructed EoR is shown in Fig.~\ref{fig:noExtreme_Full_6to20}.
The confidence regions are stretched with the increase of $z_{\rm end}$, because $\tau$ is an integral $\int x_{\rm e}^{\rm eff}\, n_{\rm H}\, {\rm d}t$ and the Planck data are more sensitive to $\tau$ than the detailed reionization process~\cite{Adam:2016hgk}.

\begin{figure*}
\begin{center}
\includegraphics[width=6in,height=5in]{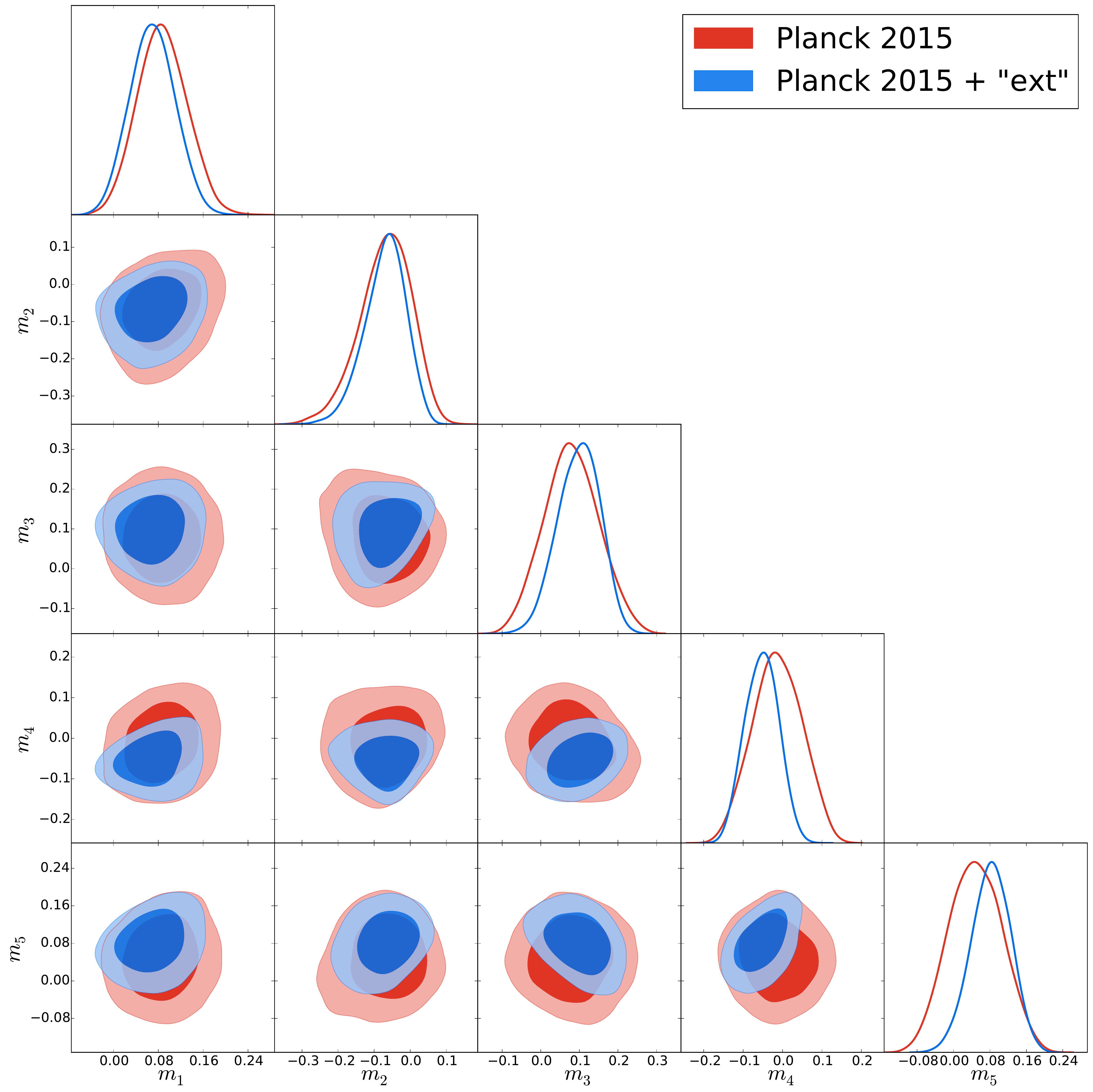}
\caption{Marginalized 2D contours ($68\%$ and $95\%$ C.L.) and posterior distributions for $m_\mu$, derived from Planck 2015 (red) and Planck 2015 + ext data (blue).}
\label{fig:joint2_tri}
\end{center}
\end{figure*}

Table~\ref{tab:parameters} summarizes the constraints on the EoR and other cosmological parameters from the Planck 2015 and $\xHI$ data.
Bounds on parameters are nearly unchanged between different models,
except the parameters of detailed reionization, the optical depth $\tau$, the degenerated parameter $A_{\rm s}$ and the rms matter fluctuations today in linear theory $\sigma_8$.
The amplitude of the primordial spectrum of scalar perturbations $A_{\rm s}$ degenerates with optical depth $\tau$ in the form $A_{\rm s} e^{-2\tau}$ on the small scale~\cite{Ade:2013zuv}, which means that a large $\tau$ leads to a large $A_s$ and $\sigma_8$.
In the PCA model with a redshift interval of $6<z<30$, the marginalized 2D contours ($68\%$ and $95\%$ C.L.) and posterior distributions for $m_\mu$ derived from Planck 2015 + ext and Planck 2015 are shown in Fig.~\ref{fig:joint2_tri}.
The ext $\xHI$ dataset is consistent with Planck 2015 data.
Constraints on the amplitudes of principal components $m_\mu$ are significantly improved in the joint analysis of Planck 2015 and ext data.

\begin{table*}[!htbp]
\renewcommand\arraystretch{1.2}
\begin{center}
\caption{Mean values and marginalized $68\%$ C.L. for reionization parameters and other cosmological parameters.}
\begin{tabular}{|>{\dtabsize}c|>{\dtabsize}c|>{\dtabsize}c|>{\dtabsize}c|}
\hline
\hline
\multirow{2}{*}{Model}   &\multicolumn{2}{>{\dtabsize}c|}{Planck 2015 + ext}  &Planck 2015 + full\\ \cline{2-4}
                            &PCA $6<z<30$        &PCA $6<z<20$                   &Instant \\
\hline
$\Omega_{\mathrm{b}}h^2$        &$0.02233\pm0.00023$     &$0.02227\pm0.00022$    &$0.02225\pm0.00023$  \\
$\Omega_{\mathrm{c}}h^2$        &$0.1187\pm0.0021$       &$0.1192\pm0.0021$      &$0.1195\pm0.0022$    \\
$100 \theta_{\mathrm{MC}}$      &$1.04102\pm0.00047$     &$1.04095\pm0.00046$    &$1.04090\pm0.00048$  \\
$\tau$                          &$0.110\pm0.014$         &$0.098\pm0.013$        &$0.083^{+0.021+0.039}_{-0.038-0.040}$      \\
$n_\mathrm{s}$                  &$0.9691\pm0.0062$       &$0.9674\pm0.0060$      &$0.9661\pm0.0062$    \\
$\ln(10^{10} A_\mathrm{s})$     &$3.151\pm0.026$         &$3.128\pm0.024$        &$3.099\pm0.044$      \\
$H_{\mathrm{0}}$\,(km\,s$^{-1}$\,Mpc$^{-1}$)      &$67.82\pm0.93$         &$67.58\pm0.94$    &$67.43\pm0.99$   \\
$\sigma_\mathrm{8}$             &$0.852\pm0.012$         &$0.844\pm0.011$        &$0.833\pm0.017$      \\
\rm{Age\,(Gyr)}                   &$13.792\pm0.037$        &$13.802\pm0.037$       &$13.807\pm0.017$     \\
$\Omega_\mathrm{\Lambda}$       &$0.692\pm0.013$         &$0.689\pm0.013$        &$0.686\pm0.014$      \\
$\Omega_\mathrm{m}$             &$0.308\pm0.013$         &$0.311\pm0.013$        &$0.314\pm0.014$      \\
$m_1$                           &$0.070\pm0.039$         &$0.204\pm0.073$        & Not applicable                \\
$m_2$                           &$-0.070\pm0.056$        &$-0.124\pm0.083$       & Not applicable                 \\
$m_3$                           &$0.098\pm0.053$         &$0.120\pm0.069$        & Not applicable                 \\
$m_4$                           &$-0.052\pm0.041$        &$-0.040\pm0.068$       & Not applicable                 \\
$m_5$                           &$0.082\pm0.042$         &$0.031\pm0.061$        & Not applicable                 \\
$z_{\rm re}$                        & Not applicable                & Not applicable            &$10.29^{+1.97+3.30}_{-3.46-3.62}$       \\
$\Delta z$                      & Not applicable           &  Not applicable              &$0.87^{+0.37+0.49}_{-0.50-0.69}$        \\
\hline
\end{tabular}
\label{tab:parameters}
\end{center}
\end{table*}

\section{Discussion and Conclusions}
\label{s:conclusion}
We have derived constraints on the cosmic reionization history using Planck temperature and low-$\ell$ polarization power spectra together with the neutral hydrogen fraction data in the $\Lambda$CDM model.
We studied the commonly adopted $\tanh$ parametrization and the PCA reionization model.
It gives unphysical results if we use the combined Planck 2015 data and the ext $\xHI$ dataset to constrain the instant model.
Meanwhile, our results show significant tension after adding the full $\xHI$ dataset in the instant model.
We may infer that the assumed instant model is oversimplified when the neutral hydrogen fraction data are included.

The PCA model is introduced to eliminate the model-dependent bias.
In the PCA model, the reconstructed $\xHI$ is consistent with $\xHI$ data.
Constraints on the low-redshift ($z\lesssim 10$) cosmic reionization history are significantly improved with the help of $\xHI$ data; nevertheless, we find that the low-redshift $\xHI$ data are nearly unhelpful for the high-redshift ($z\gtrsim 10$) constraints on $\xHI$ when combined with Planck 2015 data.
From the reconstructed reionization history, both in the case of redshift ranging from 6 to 30 and 6 to 20, we find that the Universe began to reionize at redshift no later than $z=10$ at $95\%$ C.L. Quantitatively, we derive the constraints on $\xHI$ at $z=9.75$ for both $6<z<20$ and $6<z<30$ redshift range reconstruction, and we find
\begin{eqnarray}
\xHI(z=9.75)=0.69^{+0.30}_{-0.32},
\end{eqnarray}
for $6<z<20$ reconstruction, and
\begin{eqnarray}
\xHI(z=9.75)=0.76^{+0.22}_{-0.27},
\end{eqnarray}
for $6<z<30$ reconstruction.

In the PCA model, the mean value of reionization optical depth is higher than but consistent with that obtained in the instant model.
As is shown in Fig.~\ref{fig:noExtreme_Full}, lacking direct measurements on the reionization at high redshift, constraints on the EoR are strengthened at low redshift $z\lesssim 10$ but remain nearly unchanged at high redshift $z\gtrsim 10$ by means of Planck 2015 and the $\xHI$ data.
The high-redshift EoR is only constrained by Planck 2015 data, which puts the upper limits on $x_{\rm e}$ (the lower limits on $\xHI$).
The uncertainty of $x_{\rm e}$ in the high-redshift epoch leads to a higher optical depth.
The current data are incapable of constraining the high-redshift ($z\gtrsim 10$) cosmic reionization history model independently.
Recently, Bowman {\it et al}.~\cite{Bowman:2018yin} reported an absorption profile in the sky-averaged radio spectrum of the 21-cm signal detected with the Experiment to Detect the Global Epoch of Reionization Signature (EDGES) low-band instruments.
Experiments using interferometric arrays (e.g. LOFAR~\cite{Zaroubi12}, MWA~\cite{Dillon14}, PAPER~\cite{Ali15,Pober15}, HERA~\cite{Liu16} and SKA~\cite{Mesinger15}) aimed at measuring the 21-cm signal from neutral hydrogen during the EoR have made progress.
These future experiments will probe the reionization at high redshift directly and determine the reioniation process eventually, which will also break the degeneracy between the reionization optical depth and other cosmological parameters such as the amplitude of the power spectrum of primordial scalar perturbations and neutrino masses~\cite{Liu:2015txa}.
\begin{acknowledgments}
Our numerical analysis was performed on the ``Era" of Supercomputing Center, Computer Network Information Center of Chinese Academy of Sciences.
Y.Z.M. is supported by the National Research Foundation of South Africa with Grant No.105925.
Z.K.G. is supported by the National Natural Science Foundation of China Grants No.~11690021, No.~11575272 and No.~11335012.
R.G.C. is supported by the National Natural Science Foundation of China Grants No.~11690022, No.~11435006 and No.~11647601; by the Strategic Priority Research Program of CAS Grant No.~XDB23030100; and by the Key Research Program of Frontier Sciences of CAS.

\end{acknowledgments}
\bibliography{Reion_ref}
\end{document}